\documentclass[prl,reprint,amsmath,amssymb,aps,preprintnumbers,showpacs]{revtex4-1}
\usepackage[colorlinks=true,linkcolor=black, citecolor=black,urlcolor=black]{hyperref} 
\usepackage{amstext,amsmath,amssymb,amsfonts,mathrsfs}   
\usepackage{graphicx}  
\usepackage{braket} 
\usepackage{bbm} 
\usepackage{tikz} 
\usetikzlibrary{calc}
\usetikzlibrary{patterns}
\usetikzlibrary{arrows}
\usepackage{comment}
\usetikzlibrary{plotmarks,decorations, decorations.pathmorphing}

\definecolor{dgreen}{rgb}{0,0.7,0.2}
\tikzstyle dynkin node=[very thick,shape=circle,draw,inner sep=0pt,minimum size=5mm]
\tikzstyle dynkin line=[very thick]
\tikzstyle inverse line=[gray,line width=1.46pt,line cap=round, dash pattern=on 0pt off 2\pgflinewidth]
\tikzstyle red phase=[red,thick,decoration={snake,amplitude=0.1mm,segment length=1.6mm},decorate]
\tikzstyle blue phase=[blue,thick,decoration={snake,amplitude=0.1mm,segment length=0.9mm},decorate]
\tikzstyle green phase=[dgreen,thick,decoration={snake,amplitude=0.1mm,segment length=0.9mm},decorate]
\tikzstyle brown phase=[brown,thick,decoration={snake,amplitude=0.1mm,segment length=0.9mm},decorate]
\tikzstyle arrow=[thick,rounded corners=18pt,-latex]
\tikzstyle box=[draw,rounded corners,outer sep=4pt]
\tikzstyle B node=[outer sep=0pt]
\tikzstyle Q node=[inner sep=1pt,outer sep=0pt]
\pgfdeclarelayer{background}
\pgfdeclarelayer{foreground}
\pgfsetlayers{background,main,foreground}

\newcommand{\dircoup}{\mathbf{t}}
\newcommand{\invcoup}{\mathbf{u}}
\newcommand{\auxvar}{v}

\newcommand{\AdS}{\text{AdS}}
\renewcommand{\S}{\text{S}}
\newcommand{\T}{\text{T}}
\newcommand{\CFT}{\text{CFT}}

\newcommand{\su}{\mathfrak{su}}
\newcommand{\psu}{\mathfrak{psu}}

\newcommand{\sL}{\mbox{\tiny L}}
\newcommand{\sR}{\mbox{\tiny R}}

\newcommand{\alg}[1]{\mathfrak{#1}}

\begin{document}
\preprint{Imperial-TP-RB-2016-02}

\title{On the spectrum of \texorpdfstring{$\AdS_3\times\S^3\times\T^4$}{AdS(3)xS**3xT**4} strings with Ramond-Ramond flux}
\author{Riccardo Borsato$^{1}$}
\email{r.borsato@imperial.ac.uk}
\author{Olof Ohlsson Sax$^{2}$}
\email{olof.ohlsson.sax@nordita.org}
\author{Alessandro Sfondrini$^{3}$}
\email{sfondria@itp.phys.ethz.ch}
\author{Bogdan Stefa{\'n}ski, jr.$^{4}$}
\email{Bogdan.Stefanski.1@city.ac.uk}

 \affiliation{$^1$ The Blackett Laboratory, Imperial College, London SW7 2AZ, United Kingdom\\
${}^2$ Nordita, Stockholm University and KTH Royal Institute of Technology, 
Roslagstullsbacken 23, SE-106 91 Stockholm, Sweden\\
${}^3$ Institut f\"ur Theoretische Physik, ETH Z\"urich,
Wolfgang-Pauli-Str.~27, CH-8093 Z\"urich, Switzerland\\
${}^4$ Centre for Mathematical Science, City University London,
Northampton Square, London EC1V 0HB, United Kingdom}
\date{\today}

\begin{abstract}
We analyze the spectrum of perturbative closed strings on $\AdS_3\times\S^3\times\T^4$ with Ramond-Ramond flux using integrable methods. By solving the crossing equations we determine the massless and mixed-mass dressing factors of the worldsheet S matrix and derive the Bethe equations. Using these, we construct the underlying integrable spin chain and show that it reproduces the reducible spin chain conjectured at weak coupling in arXiv:1211.1952. 
We find that the string-theory massless modes are described by gapless excitations of the spin chain. The resulting degeneracy of vacua matches precisely the protected supergravity spectrum found by de~Boer.
\end{abstract}


\maketitle

\section{Introduction}
Integrability is a powerful tool for computing generic non-protected quantities in certain gauge/string correspondences at the planar level,  which has significantly advanced our understanding of holography. It was originally discovered in  the maximally supersymmetric $\AdS_5/\CFT_4$ correspondence, see Refs.~\cite{Arutyunov:2009ga,Beisert:2010jr} for reviews, and more recently it was found to underlie $\AdS_3/\CFT_2$~\cite{David:2008yk,Babichenko:2009dk,Sundin:2012gc,Cagnazzo:2012se,Borsato:2012ud,Wulff:2014kja,Borsato:2014exa,Lloyd:2014bsa,Borsato:2015mma}, see also the review~\cite{Sfondrini:2014via}.
A quantitative handle on this duality is important as $\AdS_3/\CFT_2$ preserves half the supersymmetry of $\AdS_5/\CFT_4$, giving rise to a much richer holographic model. In fact, there are two such classes of integrable $\AdS_3$ string backgrounds: $\AdS_3\times\S^3\times\T^4$ and $\AdS_3\times\S^3\times\S^3\times \S^1$. Both can be supported by a mixture of Ramond-Ramond (RR) and Neveu-Schwarz-Neveu-Schwarz (NSNS) flux and contain a number of moduli in addition to the 't Hooft coupling. Moreover, $\AdS_3/\CFT_2$ is perhaps the first example of holography~\cite{Brown:1986nw}, has an underlying Virasoro algebra, and simple black hole solutions~\cite{Banados:1992wn}.

The present letter concerns the pure-RR $\AdS_3\times\S^3\times\T^4$ string background.
This arises as the near-horizon limit of~D1 and~D5 branes. This D1/D5 system is closely related to the moduli space of instantons~\cite{Witten:1994tz,Douglas:1996uz,Dijkgraaf:1998gf} and played a key role in string-theoretic black hole microstate counting~\cite{Strominger:1996sh}. The dual $\CFT_2$ is the infra-red fixed point of a gauge theory with both fundamental and adjoint fields. It has sixteen supercharges, giving rise to a $(4,4)$ superconformal symmetry whose global part is $\psu(1,1|2)_{\sL} \oplus \psu(1,1|2)_{\sR}$~\cite{Maldacena:1997re,Elitzur:1998mm,Larsen:1999uk,Seiberg:1999xz}.
Stringy S-duality maps the pure-RR background arising from the D1/D5 system to a pure-NSNS one. This in turn can be analyzed with worldsheet $\CFT$ techniques~\cite{Giveon:1998ns, Maldacena:2000hw}.
However, S-duality is non-planar and non-perturbative. Hence, to unravel the $\AdS_3/\CFT_2$ duality in the planar limit, one should directly tackle the RR background. It is in this setting that integrable methods are particularly useful.

The $\AdS_3\times\S^3\times\T^4$ background is classically integrable~\cite{Babichenko:2009dk,Sundin:2012gc}. Integrability should then manifest itself as factorised worldsheet scattering when the theory is quantized in light-cone gauge. However, a new feature of $\AdS_3/\CFT_2$ backgrounds is the presence of elementary {\em massless} string excitations---in the case of $\AdS_3\times\S^3\times\T^4$, the modes on the torus and their superpartners.
These had been identified as a potential challenge for integrability~\cite{Babichenko:2009dk}, especially given the subtleties of massless integrable scattering~\cite{Zamolodchikov:1992zr, Fendley:1993jh}.
Recently though, using symmetry considerations, an exact integrable worldsheet S~matrix was constructed~\cite{Borsato:2013qpa,Borsato:2014exa,Borsato:2014hja}, which incorporates massive \emph{and} massless modes~%
\footnote{%
Similar S~matrices have been also found for the $\AdS_3\times\S^3\times\S^3\times \S^1$ background as well as for mixed RR/NSNS-flux backgrounds ~\cite{Lloyd:2014bsa,Borsato:2015mma}.}.

Finding the S~matrix from the symmetries of the gauge-fixed string theory always leaves undetermined some scalar ``dressing'' factors, which are further restricted by crossing symmetry~\cite{Zamolodchikov:1978xm,Janik:2006dc}.
For $\AdS_3\times\S^3\times\T^4$ there are four such independent factors; their crossing equations were found in Refs.~\cite{Borsato:2013qpa,Borsato:2014hja}. While solutions for the two factors involving only massive excitations had already appeared in the literature~\cite{Borsato:2013hoa}, the massless and mixed-mass factors remained undetermined. 
This was because the analytic structure of the non-relativistic massless modes is a completely novel feature of $\AdS_3/\CFT_2$ that could not be deduced from $\AdS_5/\CFT_4$ integrable holography.

In this letter we solve the crossing equations for the massless and mixed-mass dressing factors by working out the analytic properties of massless excitations and the related Riemann-Hilbert problem. By diagonalizing the complete S~matrix we then find the Bethe equations for the spectrum of massless and massive excitations of the closed string. As an important check of our construction, we explicitly show how these equations have $\psu(1,1|2)^2$ symmetry.

We derive an integrable spin chain whose spectrum is encoded in the Bethe equations, and show that this agrees with the reducible spin chain originally conjectured by assuming the preservation of integrability in a massless limit  at weak coupling~\cite{Sax:2012jv}. We find that the massless modes on the worldsheet correspond to gapless excitations of the spin chain, leading to a degeneracy of the vacuum. Anticipating the result of an upcoming paper~\cite{future:details2}, we discuss how this degeneracy matches the  protected supergravity spectrum found by de Boer~\cite{deBoer:1998ip}. We believe this constitutes a strong test of our results. Some more technical details of our analysis will also be presented elsewhere~\cite{future:details1,future:details2}.

\section{Crossing and minimal solution}
The symmetries of $\AdS_3\times\S^3\times\T^4$ strings in light-cone gauge determine the dispersion relation~\cite{Borsato:2013qpa,Borsato:2014exa,Borsato:2014hja}
\begin{equation}
E(p)=\sqrt{m^2+4h^2\sin^2\tfrac{p}{2}},\quad m=\pm1,0,
\end{equation}
where $h$ is the coupling constant.
 Symmetry also fixes the two-particle integrable S~matrix $\mathbf{S}_{12}=\mathbf{S}(p_1,p_2)$~\cite{Borsato:2014exa,Borsato:2014hja} up to four dressing factors. Scattering two massive excitations gives prefactors $\sigma^{\bullet\bullet}_{12}$ or $\widetilde{\sigma}^{\bullet\bullet}_{12}$, depending on whether $m_1=m_2$ or $m_1=-m_2$. Scattering one massless and one massive excitation yields $\sigma^{\circ\bullet}_{12}$, while two massless modes give~$\sigma^{\circ\circ}_{12}$. These factors are constrained by physical and braiding unitarity and are pure phases~\cite{Borsato:2014hja}.
 \begin{figure}[t!]
   \centering
   \includegraphics{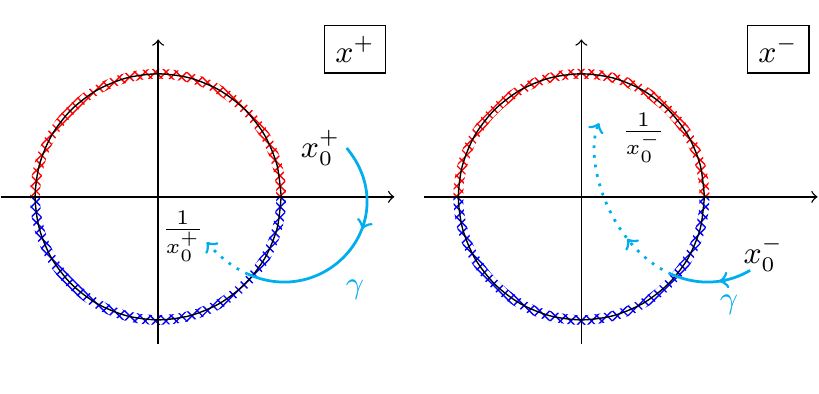}
  \caption{The crossing transformation for massive particles. Physical particles have $|x^\pm|>1$ above/below the real line. Crossing requires going through the unit circle where the dressing factors have cuts.}
  \label{fig:crossingmassive}
\end{figure}

The massive dressing factors were constructed in Ref.~\cite{Borsato:2013hoa}. The massive dispersion relation is uniformized by introducing Zhukovski variables $x^\pm$ \cite{Beisert:2004hm}, so that $E_p=\tfrac{ih}{2}(x^-_p-1/x^-_p-x^+_p+1/x^+_p)$.
The crossing transformation gives~\cite{Janik:2006dc}
\begin{equation}
\label{eq:massivecrossing}
p\to\bar{p}=-p,\ \ E_p\to E_{\bar{p}}=-E_p,\ \ x_p^\pm\to x_{\bar{p}}^\pm= \frac{1}{x_p^\pm},
\end{equation}
see Fig.~\ref{fig:crossingmassive}.  The massive crossing equations~\cite{Borsato:2013qpa} are solved 
in terms of the Beisert-Eden-Staudacher (BES) phase~\cite{Beisert:2006ib,Beisert:2006ez}, the Hern\'andez-L\'opez (HL) phase~\cite{Hernandez:2006tk} and a novel function ~$\sigma^-$ which distinguishes the two massive phases, $\sigma^{\bullet\bullet}/\widetilde{\sigma}^{\bullet\bullet}= \sigma^-$ ~\cite{Borsato:2013hoa}. This matches several perturbative computations~\cite{Rughoonauth:2012qd, Beccaria:2012kb, Sundin:2013uca,Abbott:2013mpa, Engelund:2013fja, Roiban:2014cia, Bianchi:2014rfa, Sundin:2014ema, Sundin:2015uva,  Abbott:2015pps}.

\begin{figure}[t!]
  \centering
  \includegraphics{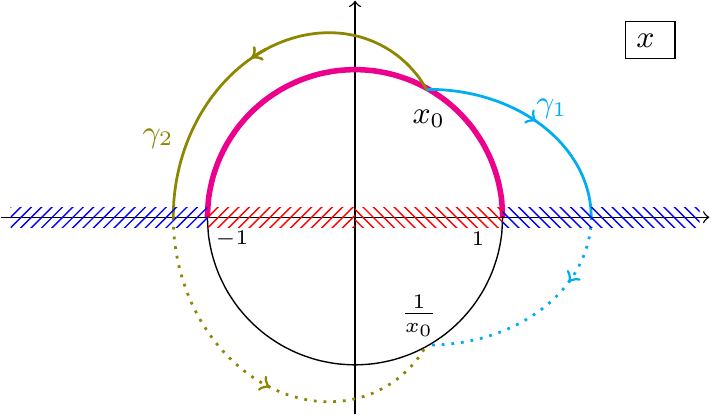}
  \caption{In the massless $x$-plane, the physical region is the upper half-circle (magenta line). We expect the dressing factors to have branch cuts where the real part of $E_p$ changes sign, \textit{i.e.}\ on the real line. Crossing sends $x_0\to1/x_0$ through such cuts.}
  \label{fig:crossingX}
\end{figure}

While the crossing transformation for massive excitations is well-understood~\cite{Janik:2006dc,Arutyunov:2009ga,Borsato:2013hoa}, particles with $m=0$ present entirely new features. Introducing the gapless Zhukovski variables~%
\footnote{
  These can be obtained from the usual Zhukovski variables as $(x_p)^{\pm 1}=\lim_{m\to0}x^{\pm}_p$ \cite{Borsato:2014exa,Borsato:2014hja}. Note that in the massless limit $x^+$ and $x^-$ are related, $\lim_{m\to0}x^{+}_p\,x^{-}_p=1$.
}
$x_p=e^{ip/2}\text{sgn}[\sin\tfrac{p}{2}]$, the dispersion relation uniformises, $E_p=-i  h(x_p-1/x_p)$. Crossing reads similarly to~\eqref{eq:massivecrossing}, with $x_p\to x_{\bar{p}}=1/x_{p}$. A crucial difference is that the physical region for $x_p$ lies on the unit circle, see Fig.~\ref{fig:crossingX}.
Crossing symmetry requires the dressing factors to satisfy~\cite{Borsato:2013qpa}
\begin{equation}
\label{eq:crossingeq}
\begin{aligned}
&\sigma^{\circ\circ}(\bar{p}_1,p_2)^2 \sigma^{\circ\circ}(p_1,p_2)^2= F(w_1-w_2)\,f(x_1,x_2)^2,\\
&\sigma^{\circ\bullet}(\bar{p}_1,p_2)^2 \sigma^{\circ\bullet}(p_1,p_2)^2=\frac{f(x_1,x^+_2)}{f(x_1,x^-_2)}\,,
\end{aligned}
\end{equation}
with $F(w)=1+i/w$ and $f(x,y)=\tfrac{xy-1}{x-y}$.

Let us firstly consider $\sigma^{\circ\circ}$. Its crossing equation involves the rapidity~$w_p$, which emerges from an $\su(2)$ invariance of~${\bf S}_{12}$, and satisfies~$w_{\bar{p}}=w_p+i$. It is straightforward to construct non-trivial solutions for the $w$-dependent part of crossing 
\footnote{%
The simplest solution is taking $w_p$ to be a massless limit of Janik's rapidity~$z_p$~\cite{Janik:2006dc} and the $w$-dependent part of $\sigma^{\circ\circ}$ to be the dressing factor of the $SU(2)$ chiral Gross-Neveu model~\cite{Gross:1974jv,Berg:1977dp}. We thank Alessandro Torrielli for elucidating this point.}. However, none is consistent with perturbation theory~\cite{Sundin:2015uva,future:LinusAndPer}. For this reason we conjecture that the $\su(2)$ S~matrix $\mathbf{S}^{\su(2)}_{12}$ of Ref.~\cite{Borsato:2014exa,Borsato:2014hja} trivialises together with its dressing factor, which amounts to taking $w\to\infty$. 

\begin{figure}[t!]
\centering
\includegraphics{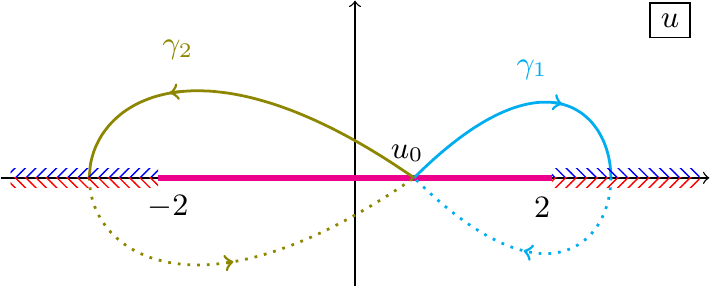}
  \caption{The $u$-plane. The thick magenta line indicates real momenta, and the dashed lines indicate the discontinuities of the energy. Crossing sends $u_0$ to its image on the next sheet.}
  \label{fig:crossingU}
\end{figure}

By iterating the crossing transformation twice, $x_p$ goes to itself, $x(\bar{\bar{p}})=x(p)$. However, for $\sigma^{\circ\circ}$ we find $\sigma^{\circ\circ}(\overline{\overline{p}}_1,p_2)\neq\sigma^{\circ\circ}(p_1,p_2)$. This implies that the simplest solution of crossing must have cuts in the $x$-plane, \textit{cf.}\ Fig.~\ref{fig:crossingX}. To construct such a minimal solution
\footnote{Crossing symmetry only determines the dressing factors up to ``CDD factors''~\cite{Castillejo:1955ed}.}
for Eqn.~\eqref{eq:crossingeq} we introduce the variable~$u=x+1/x$. The branch-cuts of the energy are mapped to real $u$ with $|u|>2$, and the crossing transformation takes $u_0$ to itself as in Fig.~\ref{fig:crossingU}. The logarithm of the crossing equation can be analytically continued so that $u_0$ is just above the cut. This yields a Riemann-Hilbert problem for $\theta^{\circ\circ}=-i\log \sigma^{\circ\circ}$
\begin{equation}
\theta^{\circ\circ}(u_1+i0,u_2)+\theta^{\circ\circ}(u_1-i0,u_2)=-i\log f(x_1,x_2), 
\end{equation}
which can be solved by standard techniques~\cite{future:details1}. Going back to the $x$-plane and setting 
\begin{equation}
\label{eq:minimalsolution}
\theta_{12}^{(\pm)} \!=\! \pm\!\!\!\!\!\int\limits_{-1\pm i0}^{+1\pm i0}\!\!\!\!\frac{dz}{2\pi}g_{\mp}(z,x_2)\partial_z g_{\mp}(z,x_1)\mp\frac{i}{2} g_{\pm}(x^{\pm1}_1\!\!,x^{\pm1}_2),
\end{equation}
where $g_\pm(z,x)=\log[\pm i(x-z)]-\log[\pm i(x-1/z)]$, we  have $\theta^{\circ\circ}=\theta^{(+)}+\theta^{(-)}$.

Following Ref.~\cite{Arutyunov:2004vx} we rewrite the phase $\theta^{\circ\circ}$ as a series over conserved charges
\begin{equation}
\label{eq:crsexpansion}
\theta^{\circ\circ}_{12} =\sum_{r,s}c^{\circ\circ}_{r,s} (\mathcal{Q}_{r}(x_1)\mathcal{Q}_{s}(x_2)-\mathcal{Q}_{s}(x_1)\mathcal{Q}_{r}(x_2)),
\end{equation}
where for gapless modes~$\mathcal{Q}_{r+1}(x)=\frac{ih}{r}(x^{-r}-x^{r})$~\footnote{This is simply the massless limit of the usual charges~\cite{Arutyunov:2004vx}.}.
The coefficients $c_{r,s}^{\circ\circ}$ match those obtained at one-loop in the worldsheet calculation of Ref.~\cite{Sundin:2015uva} and as noted there coincide with those of the HL phase \cite{Hernandez:2006tk}. As ours is an \emph{all-loop} solution, this suggests a drastic simplification of crossing when going from massive to massless kinematics.

To see such a simplification, we can formally take the massless limit in the crossing equations of $\sigma^{\bullet\bullet},\, \widetilde{\sigma}^{\bullet\bullet}$. Then the phases can be taken to be equal and each must solve the massless crossing equation, $\sigma^{\bullet\bullet}(\bar{p}_1,p_2)\sigma^{\bullet\bullet}(p_1,p_2) =f(p_1,p_2)$.
Moreover, we can take a massless limit on the \emph{solutions} of the crossing equations. By working order by order in an asymptotic large-$h$ expansion~\cite{Beisert:2006ib,Vieira:2010kb,Borsato:2013hoa} one can show that all terms beyond HL order vanish when evaluating~$\sigma^{\bullet\bullet},\, \widetilde{\sigma}^{\bullet\bullet}$ for massless kinematics, and that in that limit~$\sigma^{-}\to1$ so that the two phases coincide~\cite{future:details1}. Therefore we expect that the minimal solution~\eqref{eq:minimalsolution} captures the relevant physics in the massless sector despite its apparent simplicity.

The minimal solution for $\sigma^{\circ\bullet},$ can be found by similar considerations~\cite{future:details1}. The phase can be expanded as in Eq.~\eqref{eq:crsexpansion} with appropriate massive/massless kinematics for the charges~$\mathcal{Q}_{r}$. One can then show that the coefficients~$c_{r,s}^{\circ\bullet},$ equal~$c_{r,s}^{\circ\circ}$, and that this solution too can be thought of as limits of the massive ones~\cite{future:details1}.

\section{Bethe equations}

\begin{figure}
  \centering
  \includegraphics{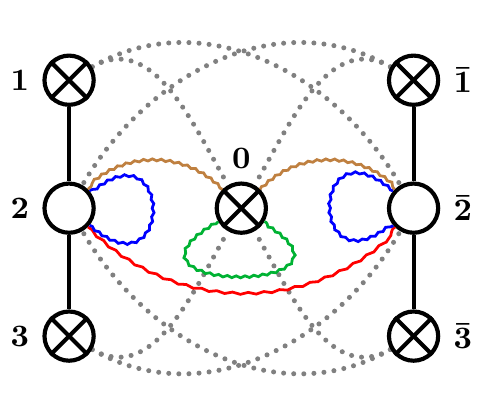} 

  \caption{We represent the Bethe equations with two copies of the Dynkin diagram for $\psu(1,1|2)$ supplemented by one node for massless fermions. We use solid lines for Dynkin links, and dotted lines for the other interactions between auxiliary nodes and momentum carrying ones. Blue and red wavy links are used for dressing phases of the massive sector $\sigma^{\bullet\bullet}$ and $\widetilde{\sigma}^{\bullet\bullet}$ respectively. Brown wavy links represent the dressing phase $\sigma^{\circ\bullet}$, while the green one represents $\sigma^{\circ\circ}$.}
  \label{fig:bethe-equations}
\end{figure}
Imposing that the wave-function of closed strings is periodic on a circle of length~$L$ we find the Bethe equations.
Together with level-matching $\prod_k e^{ip_k}=1$, they give quantisation conditions for momenta $p_k$ of the worldsheet excitations. 
In Fig.~\ref{fig:bethe-equations} we depict the Bethe equations by associating a node to each set of roots, and by linking the nodes with lines representing the various interactions.
Given the complexity of the S~matrix we use a diagonalisation procedure, meaning that together with the momenta $p_k$ associated to nodes $K\in \mathbf{m}=\{2,\bar{2},0\}$ we also have auxiliary roots $\auxvar_{K,k}$ related to nodes $K\in \mathbf{a}=\{1,3,\bar{1},\bar{3}\}$.
These two sets of variables satisfy respectively the following two Bethe equations
\begin{eqnarray}
\nonumber\displaystyle
&e^{i p_k L} = \displaystyle\prod_{J\in \mathbf{m}} \prod_{\substack{j = 1\\j \neq k}}^{N_J}S_{KJ}(x^\pm_k,x^\pm_j)
 \prod_{J\in \mathbf{a}}\prod_{j=1}^{N_J} S_{KJ}(x^\pm_k,\auxvar_{J,j}),\\
\displaystyle
&1=\displaystyle\prod_{J\in \mathbf{m}} \prod_{j=1}^{N_J} S_{KJ}(\auxvar_{K,k},x^\pm_j) ,
\end{eqnarray}
where $x_k^{\pm} = x^{\pm}(p_k)$~\footnote{%
It is convenient to formally use~$x^\pm$ for massless modes, meaning $x^+=x,$ $x^-=1/x$.
}.
The factors $S_{KJ}$ satisfy $S_{JK}=S_{KJ}^{-1}$ as a consequence of unitarity.
The momentum-carrying nodes in $\mathbf{m}$ correspond to the highest weight states of each module.

Left-massive excitations on S$^3$ and right-massive ones on AdS$_3$ correspond to nodes $2$ and $\bar{2}$, respectively. They were denoted by $Y^{\sL}$, $Z^{\sR}$ in Refs.~\cite{Borsato:2014exa,Borsato:2014hja}.
Massless fermions sit at the node $0$, and transform in a doublet $\chi^{\alpha}$  of $\su(2)_{\circ}$. This auxiliary $\su(2)_{\circ}$ symmetry commutes with $\psu(1,1|2)^2$ and acts on all massless modes. All scattering processes involving these excitations are diagonal and they produce the corresponding factors $S_{KJ}$
\footnote{The factors appearing in the Bethe equations are the inverse of the corresponding scattering processes, \textit{e.g.}\ 
$S_{22}^{-1}(x^\pm_k,x^\pm_j)= {\langle Y^{\text{L}}_{j} Y^{\text{L}}_{k}|\mathbf{S} |Y^{\text{L}}_{k} Y^{\text{L}}_{j} \rangle}$.
 Here we write the results in the spin-chain frame. It was also convenient to modify the normalisation in the mixed-mass sector from that of Ref.~\cite{Borsato:2014hja}.}
\begin{equation}
\begin{aligned}
& S_{22}= \dircoup^{+-}_{-+}\, \invcoup^{+-}_{-+} (\sigma^{\bullet\bullet})^2,
\, 
&& S_{02}=   (\dircoup^{+-}_{-+})^{\frac{1}{2}}(\dircoup^{--}_{++})^{\frac{1}{2}} (\sigma^{\circ\bullet})^2,
\\
& S_{\bar{2}\bar{2}}=\dircoup^{-+}_{+-} \, \invcoup^{+-}_{-+} (\sigma^{\bullet\bullet})^2,
&& S_{0\bar{2}}=   (\invcoup^{+-}_{-+})^{\frac{1}{2}}(\invcoup^{++}_{--})^{\frac{3}{2}}     (\sigma^{\circ\bullet})^2,
\\
& S_{2\bar{2}}= \invcoup^{++}_{--} \, \invcoup^{+-}_{-+}  (\widetilde{\sigma}^{\bullet\bullet})^2,
&& S_{00}=   \dircoup^{+-}_{-+}(\sigma^{\circ\circ})^2.
\end{aligned}
\end{equation}
Above, we dropped the dependence on $(x^\pm_k,x^\pm_j)$ for brevity, and introduced the functions
\begin{equation}
\dircoup^{\mathsf{ab}}_{\mathsf{cd}}(x_k,x_j)
=\frac{x^{\mathsf{a}}_k -x^{\mathsf{b}}_j}{x^{\mathsf{c}}_k -x^{\mathsf{d}}_j},
\
\invcoup^{\mathsf{ab}}_{\mathsf{cd}} (x_k,x_j)
 =\frac{1-(x^{\mathsf{a}}_k x^{\mathsf{b}}_j)^{-1}}{1-(x^{\mathsf{c}}_k x^{\mathsf{d}}_j)^{-1}}.
\end{equation}
The auxiliary nodes in $\mathbf{a}$ correspond to supercharges which turn the excitations $Y^{\sL}$, $Z^{\sR}$ and $\chi^\alpha$ into their superpartners $\eta_{\sL}^{\ a}$, $\eta_{\sR a}$ and $T^{a\alpha}$, in the notation of Ref.~\cite{Borsato:2014exa,Borsato:2014hja}.
Scattering processes which include also these excitations are not diagonal, and the corresponding factors $S_{KJ}$ can be derived using the nesting procedure~\cite{Yang:1967bm}. We find that these nodes interact only with the momentum-carrying ones, and for $K=1,3$
\begin{equation}
\begin{aligned}
S_{2K}=  (\dircoup^{-\, \cdot}_{+\, \cdot}) ,\quad
S_{\bar{2}K}= (\invcoup^{+\, \cdot}_{-\, \cdot}) ,\quad
S_{0K}=  (\dircoup^{-\, \cdot}_{+\, \cdot}),
\end{aligned}
\end{equation}
where we use a dot to indicate that no superscript is needed on auxiliary roots.
For $K=\bar{1},\bar{3}$ one needs to swap $\dircoup$ and $\invcoup$ in the above expressions.

If we had a non-trivial S~matrix for the $\su(2)_\circ$ of massless excitations, the Bethe equations would have an additional node accompanied by the corresponding auxiliary roots.
As this is not the case, the node $0$ represents at the same time both massless fermions $\chi^1$ and $\chi^2$, which should be taken into account when enumerating the states~\cite{future:details2}.

A consistency condition for this construction is the re-emergence of the global $\psu(1,1|2)^2$ symmetry. This symmetry appears because the Bethe equations remain invariant when we add roots $x^\pm$ at infinity---corresponding to zero momentum---for nodes $2$ and $\bar{2}$, or similarly for auxiliary roots $\auxvar_{K,k}$. Following Ref.~\cite{Beisert:2005fw}, we can also read off the global charges $D_{\sL,\sR}$ and $J_{\sL,\sR}$ corresponding to the Left and Right $\alg{sl}(2)$ and $\su(2)$ subalgebras by further expanding the roots $x^\pm$ at infinity at subleading order 
\begin{equation}
\label{eq:globalcharges}
  \begin{aligned}
D_{\sL} &= \tfrac{1}{2} (L + N_1 + N_3 - \phantom{2}N_0 + \delta D),\\
D_{\sR} &= \tfrac{1}{2} (L - N_{\bar{1}} - N_{\bar{3}} + 2N_{\bar{2}} + \delta D),\\
J_{\sL} &= \tfrac{1}{2} (L + N_1 + N_3 - 2N_2 - N_0 ),\\
J_{\sR} &= \tfrac{1}{2} (L - N_{\bar{1}} -N_{\bar{3}}),
  \end{aligned}
\end{equation}
where $\delta D=ih\,\sum_{K\in\mathbf{m}} \sum_{k=1}^{N_K} \left( 1/x_k^+ - 1/x_k^-\right)$ is the anomalous dimension.

The diagram in Fig.~\ref{fig:bethe-equations} encodes the Bethe equations and, should we interpret it as a Dynkin diagram, would hint at a symmetry enhancement beyond $\psu(1,1|2)^2$. It would be interesting to explore this point further.

\section{Spin chain and protected states}

Two natural and related questions to ask are whether there is a spin chain whose spectrum is captured by the above Bethe equations and what the set of protected states is. When there are no massless excitations we get back the equations derived in Ref.~\cite{Borsato:2013qpa}. As explained there, these correspond to a homogeneous spin chain where the sites transform in identical representations of $\psu(1,1|2)^2$. For a spin chain of $J$ sites this ground state has conformal weight $(D_{\sL},D_{\sR}) = (\tfrac{J}{2},\tfrac{J}{2})$ and satisfies the $\tfrac{1}{2}$-BPS shortening condition $D_{\sL} + D_{\sR} = J_{\sL} + J_{\sR}$, corresponding to a highest weight state with weights $(\tfrac{1}{2},\tfrac{1}{2})$ at each site. 

Let us add a single massless Bethe root by setting $N_0=1$, and increase the length $L$ by one. From the level-matching constraint this excitation must have zero momentum and hence no anomalous dimension. From the global charges~\eqref{eq:globalcharges} we find that the $\tfrac{1}{2}$-BPS condition $D_{\sL} + D_{\sR} = J_{\sL} + J_{\sR}$ is still satisfied. However, the weights of the new state are $(\tfrac{J}{2},\tfrac{J+1}{2})$. Hence, we can interpret the addition of the massless Bethe root as adding a single chiral site with weights $(0,\tfrac{1}{2})$. 

In addition to the massless root, we can also add two auxiliary roots of type $1$ and $\bar{3}$. This again leads to a $\tfrac{1}{2}$-BPS state but now with weights $(\tfrac{J+1}{2},\tfrac{J}{2})$, corresponding to adding a site with weights $(\frac{1}{2},0)$. As discussed in the previous section, each massless root corresponds to a doublet of $\su(2)_{\circ}$. Altogether, we find four fermionic zero modes stemming from the massless excitations.

Anticipating a result of Ref.~\cite{future:details2}, let us see how these zero modes can be used to construct protected operators of arbitrary length. For states with several massless excitations we need to solve the Bethe equations to determine the location of the roots. In order to find the protected states we note that the basic massless excitations discussed above are fermionic. This means that each of the four modes can appear at most once for a given momentum. At the same time, a \emph{non-zero} momentum would lead to an anomalous dimension. As a result, we are left with a tower of sixteen $\tfrac{1}{2}$-BPS states starting from a given ground state not containing any massless excitations. The conformal weights and multiplicities of these states can be conveniently organised in the following diamond
\begin{equation*}
  \begin{tikzpicture}
\newlength{\hordist}
\setlength{\hordist}{1.5cm}
\newlength{\vertdist}
\setlength{\vertdist}{0.7cm}

    \node at (0,0) {$(\tfrac{J}{2},\tfrac{J}{2})$};

    \node at (-1\hordist,-1\vertdist) {$(\tfrac{J}{2}+\tfrac{1}{2},\tfrac{J}{2})^{\oplus 2}$};
    \node at (+1\hordist,-1\vertdist) {$(\tfrac{J}{2},\tfrac{J}{2}+\tfrac{1}{2})^{\oplus 2}$};

    \node at (-2\hordist,-2\vertdist) {$(\tfrac{J}{2}+1,\tfrac{J}{2})$};
    \node at (0,-2\vertdist) {$(\tfrac{J}{2}+\tfrac{1}{2},\tfrac{J}{2}+\tfrac{1}{2})^{\oplus 4}$};
    \node at (+2\hordist,-2\vertdist) {$(\tfrac{J}{2},\tfrac{J}{2}+1)$};

    \node at (-1\hordist,-3\vertdist) {$(\tfrac{J}{2}+1,\tfrac{J}{2}+\tfrac{1}{2})^{\oplus 2}$};
    \node at (+1\hordist,-3\vertdist) {$(\tfrac{J}{2}+\tfrac{1}{2},\tfrac{J}{2}+1)^{\oplus 2}$};

    \node at (0,-4\vertdist) {$(\tfrac{J}{2}+1,\tfrac{J}{2}+1)$};
  \end{tikzpicture}
\end{equation*}
where the eight states in the second and fourth row are fermionic, and the remaining eight are bosonic.
This set of $\tfrac{1}{2}$-BPS states agrees exactly with the protected supergravity spectrum for $\AdS_3 \times \S^3 \times \T^4$~\cite{deBoer:1998ip}. As a result the perturbative closed string part of the modified elliptic genus of the two models matches~\cite{future:details2}.

The above discussion further leads to an interesting picture of a spin chain that includes both massive and massless excitations. The resulting spin chain is \emph{inhomogeneous}: there are multiple short irreducible representations of $\psu(1,1|2)^2$ in which the sites can transform, with conformal weights $(\tfrac{1}{2},\tfrac{1}{2})$, $(\tfrac{1}{2},0)$ and $(0,\tfrac{1}{2})$, respectively. Moreover, the spin chain is \emph{dynamic}: energy eigenstates will be linear combinations of states with a different assignment of irreducible representations at each site. Finally, the spin-chain Hamiltonian contains \emph{length-changing} interactions~\cite{Borsato:2013qpa}. This spin-chain structure agrees with the ``reducible spin~chain'' proposed as a model incorporating massless modes in Ref.~\cite{Sax:2012jv}.

\section{Outlook}
In this letter we derived Bethe equations for the spectrum of closed string states on $\AdS_3\times\S^3\times\T^4$ with RR flux. These equations incorporate both massive and massless worldsheet excitations.  It would be important to understand the wrapping corrections of massive and massless
particles, a discussion of which was recently initiated in the present context in Ref.~\cite{Abbott:2015pps}. We determined the analytic structure of the massless modes and found solutions of the massless and mixed mass crossing equations. We then proposed a spin chain whose spectrum is encoded in the Bethe equations. We found that this spin chain corresponds to the reducible spin chain first proposed  at weak coupling in Ref.~\cite{Sax:2012jv}. In particular, the worldsheet massless modes correspond to spin-chain gapless excitations, resulting in a degeneracy of the vacuum. This degeneracy reproduces the  protected supergravity spectrum found by de Boer~\cite{deBoer:1998ip}, providing a strong test of our results~\cite{future:details2}.

Since integrable S~matrices exist for a wide variety of $\AdS_3$ backgrounds~\cite{Borsato:2014exa,Lloyd:2014bsa,Borsato:2015mma}, the construction presented here should be adapted to those cases. In particular, it would be interesting to determine the effect of NSNS flux on the spin chain and whether one may approach the Wess-Zumino-Witten point with integrable methods. Further, the derivation of an integrable spin chain for the $\AdS_3\times\S^3\times\S^3\times \S^1$ background and its vacuum degeneracy is likely to provide important clues about the enigmatic $\CFT_2$ dual of this background~\cite{Gukov:2004ym,Tong:2014yna}. Another open problem is how the finite-gap limit of the Bethe equations relates to the semi-classical analysis of~\cite{Lloyd:2013wza,Abbott:2014rca}.

It is an important question to find integrable structures on the $\CFT_2$ side of the duality. While results at the symmetric orbifold point seem negative~\cite{Pakman:2009mi}, large-$N_f$ analysis of the IR fixed point of the dual gauge theory~\cite{Sax:2014mea} has provided evidence for integrability and the reducible spin chain discussed here. It would also be interesting to relate our results to higher spin theories such as the ones recently considered in Ref.~\cite{Gaberdiel:2014cha}.

Integrable methods are beginning to shed new light on the $\AdS_3/\CFT_2$ correspondence which we hope will lead to a better understanding of this duality.

\vspace{-0.2cm}
\subsection{Acknowledgements}
We thank Gleb Arutyunov, Marco Baggio, Per Sundin, Alessandro Torrielli, Arkady Tseytlin, Linus Wulff, Kostya Zarembo for discussions and useful comments on the manuscript. Thanks to Sergey Frolov, Ben Hoare, Tom Lloyd for discussions. We are particularly grateful to M.~Baggio and A.~Torrielli for their collaboration on part of the research appearing here which will be presented and discussed in more detail elsewhere~\cite{future:details1,future:details2}.
We also thank P.~Sundin and L.~Wulff for sharing and discussing with us an early draft of Ref.~\cite{future:LinusAndPer}.
We would also like to acknowledge numerous discussions with A.~Torrielli on $\AdS_3/\CFT_2$.
R.B.\@ was supported by the ERC Advanced grant No.\@ 290456.
O.O.S.\@ was supported by ERC Advanced grant No.\@ 341222.
A.S.’s research was partially supported by the NCCR SwissMAP, funded by the Swiss National Science Foundation.
B.S.\@ acknowledges funding support from an STFC Consolidated Grant
ST/L000482/1.\@ R.B., O.O.S.\@ and B.S.\@ would like to thank ETH Zurich and all participants of the workshop \textit{All about $\AdS_3$} for providing a stimulating atmosphere where parts of this work were undertaken.\@ R.B., A.S.\@ and B.S.\@ would like to thank Nordita for hosting us during the final stages of this project.

\bibliographystyle{h-physrev}
\bibliography{refs}

\end{document}